# MODELING HUMAN INTERACTION TO DESIGN A HUMAN-COMPUTER DIALOG SYSTEM

A. Loisel, N. Chaignaud and J-Ph. Kotowicz

*LITIS Laboratory - EA 4108 - Place Emile Blondel - BP 08 - 76131 Mont-Saint-Aignan Cedex*

*{loisel, chaignaud, kotowicz}@insa-rouen.fr*

Abstract: This article presents the Cogni-CISMeF project, which aims at improving the health information search engine CISMeF, by including a conversational agent that interacts with the user in natural language. To study the cognitive processes involved during information search, a bottom-up methodology was adopted. An experiment has been set up to obtain human dialogs related to such searches. The analysis of these dialogs underlines the establishment of a common ground and accommodation effects to the user. A model of artificial agent is proposed, that guides the user by proposing examples, assistance and choices.

## 1 INTRODUCTION

CISMeF (French acronym for "Catalog and Index of French-language health resources" www.cismef.org) aims at describing and indexing the main French-language health resources in order to assist health professionals and consumers in their search for electronic information available on the Internet. To index resources, CISMeF uses four different concepts: meta-term, keyword, subheading and resource type. It contains a thematic index, including medical specialties, and an alphabetic index. Nowadays, the system includes a graphic user-interface, a query language and uses index and thesaurus to find information. However, the "extended" and the "boolean" search options increase the complexity of the interface and users are not comfortable with it.

The aim of the Cogni-CISMeF project is to improve search in CISMeF by including a conversational agent that interacts with the user in natural language. This agent leads the user in his information search by analyzing his aims and by proposing, assistance and choices. Once recognized, the user's intention is translated into queries.

In order to adapt the system to the user, we believe that the human-computer interactions shall be designed to mimic human interactions. To this end, an experiment has been set up to obtain human dialogs between a CISMeF expert and users looking for health information. These dialogs (constituting a corpus) have been analyzed to extract their discursive structure and their linguistic features in order to build a cognitive model of a conversational agent.

In this article, Section 2 describes related work on dialog systems. Section 3 details the psychological experiment we have set up and the corpus collection. The analysis of the corpus is presented in Section 4 and Section 5 describes the cognitive model that we propose, according to these results. In Section 6, conclusion and perspectives close this paper.

## 2 DIALOG SYSTEMS

Theories used by human-computer dialog systems can be classified into several categories. One possibility is to assess whether they are based on the agent intention or on social conventions.

### 2.1 Intention Based Approaches

Intention based approaches use a representation of the mental states of the artificial agent. The most famous model is BDI (Belief, Desire and Intention). which has been used both in logic (Cohen and Levesque, 1990) and planning (Allen and Perrault, 1980) settings. Its implementation is complex and its reuse is domain restricted.

## 2.2 Convention Based Approaches

To simplify, a dialog can be considered as a protocol represented by finite state automata in which transitions are the possible speech acts of the dialog. The agent has no internal representation. These approaches are rather rigid even if some of them (Sitter and Stein, 1992) use recursive automata.

Another conventional model (Lewis, 1979) consists in representing information shared during the dialog (called "common ground") in a conversational board. This theory is more descriptive than predictive and thus is difficult to integrate into a dialog system.

## 2.3 Mixed Approaches

Dialog games (Levin and Moore, 1980) are interested in social conventions between utterances. They use structures, games for which interactions are precisely described. Games are stereotypes that model a communicational situation.

The QUD (Questions Under Discussion) model, proposed by (Ginzburg, 1996) and totally implemented in the GoDiS system (Larsson, 2002), takes into account mainly the transmission of missing information. The dialog uses both a conversational board and internal representation of the agent. This approach is mainly based on the questions and their responses. Each speech act (enunciated by the user or the system) modifies the "information state" (IS), comprising a private part and a public part.

With the "grounding" theory, (Traum, 1994) proposes 5 modalities according to which an utterance is grounded: perception, contact, semantic understanding, pragmatic understanding, integration. For each modality, there are speech acts of positive (resp. negative) grounding if this modality is (resp. is not) grounded. For example, if the perception is grounded but not the semantic understanding, the system can produce a repeating of the utterance to show that it has been heard and then it can say a speech act like "not understood".

This approach is highly capable when it is added with accommodation effects (Lewis, 1979) like in GoDiS. When user utterances do not match with the current plans, the system loads a new relevant plan to this utterance. Plans can be performed in parallel.

## 3 CORPUS COLLECTION

At first, we wanted to model the reasoning of the CISMeF chief librarian, when he was searching in the CISMeF system. He was asked five questions from health professionals and his answers have been recorded. These records showed that the CISMeF chief librarian has a complete understanding of the user's intention and suggests optimal queries. However, he does not need to converse with the user to understand his inquiry. We had thus to set up a new experimentation dealing with the recording of dialogue between a CISMeF expert and a user.

The users were voluntary members of the LITIS laboratory (secretary, PhD students, researchers and teachers) who wanted to obtain responses about medical inquiries. The experts were two members of our project, trained to the CISMeF system and terminology. The experimentation took place as follows: one expert and one user were facing a computer using the advanced search interface of the system and recording all the queries with their answers in a log. The expert was in charge of conducting the search by conversing with the user and verbalizing each action, inquiry and answer. The experimentation ended when relevant documents were given to the user or when it seemed that no answer existed in the system. A textual corpus was constituted from the transcription of the twenty-one dialogues recorded.

Moreover, following this experimentation, we asked the CISMeF chief librarian to answer the users' inquiries and to verbalize his search process. The verbal occurrences were also recorded. Our aim was to obtain optimal queries to these questions using the CISMeF terminology. They provide explanations about the strategies adopted by the chief librarian.

## 4 ANALYSIS OF THE CORPUS

We have hand-analyzed the textual corpus. During the conversations, experts tried to keep control of the dialog by making the user repeat and confirm his utterances to avoid ambiguity or contestation. Many discursive tags (agreement, question, suggestion, refusal…) lead to interaction. Several iterative loops ensure the continuity of the dialog.

This analysis brings out a global structure of dialogs broken down into sub-dialogs and it allows to build a list of speech acts observed in the corpus.

## 4.1 Global Structure of Dialogs

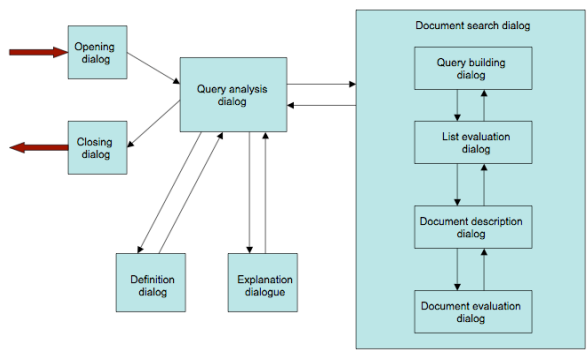

Figure 1: Links between sub-dialogs

In the dialogs, there are a lot of comings and goings between the initial query of the user and the answers of the system depending on the results. Moreover, dialogs can be divided into sub-dialogs. Figure 1 describes the possible links between sub-dialogs. A dialog always begins with an *opening sub-dialog*, which can indifferently be short or long. It consists in identifying the user, presenting the CISMeF system and negotiating the task. Then, the user can ask the expert his medical inquiry in a *querying sub-dialog*. The expert reformulates the question to be sure of the tackled themes and the meaning of the words used. The inquiry can be broken down into several other inquiries that can be a question about a definition or about explanation on the system itself. In the case of an information inquiry, the expert builds the query with the help of the user. Each term constituting the query is discussed according to the CISMeF terminology. Queries are performed and the list of documents is presented to the user. One particular document can be described. At any time, these sub-dialogs can be interrupted by precision inquiries. The dialog finishes with an *ending sub-dialog* on the initiative of the user either with a success (the documents are relevant) or with a failure.

## 4.2 Taxonomy of Speech Acts

A list of speech acts has been built according to linguistic features found into the corpus.
This taxonomy comes from (Weisser, 2003) and has been adapted to our corpus. It follows the illocutionary force of the speech acts.

*Initiative assertives*
- `Inform`: to bring information without expecting any response
(e.g. expert: "I think that the keyword "parasomny" also exists")

*Initiative directives*
- `RequestInfo`: information query
(e.g. expert: "Do you think that we can find a medical specialty?")
- `Offer`: to propose something that the interlocutor can accept or refuse
(e.g. expert: "Do you want to try with the keyword "general medicine"?")
- `RequestDirective`: the speaker expects guidelines from the interlocutor
(e.g. expert: "What is your question?")

*Reactive assertives*
- `Answer`: response to a question
(e.g. expert: "There are to many documents!")
- `Accept`: to agree with a previous utterance that is both achieved and satisfied
(e.g. user: "Yes, exactly!")
- `Refuse`: to refuse a previous utterance that is achieved but not satisfied
(e.g. user: "No, I am not interested")
- `Acknowledge`: to tell the interlocutor that his utterance is achieved
(e.g. expert: "Ok! I understood the question!")
- `WantsNothing`: to answer negatively to a `RequestDirective`
(e.g. user: "No, I do not want anything else")

*Reactive directives*
- `Confirm`: request of utterance confirmation
(e.g. expert: "You want to know the process to follow to donate an organ, don't you?")

*Declaratives*
- `Bye`: to conclude the conversation and to close the communication channel
(e.g. expert: "Bye, have a nice day!")
- `Greet`: to initiate a conversation or to pursue it after a break
(e.g. expert: "Hello, what is your question?")

*Promissives*
- `InformIntent`: to specify to the interlocutor what we are about to do
(e.g. expert: "Well, let's see if we can find something about it")

Some of these acts are explicit « grounding » acts: `Accept, Acknowledge, WantsNothing, Confirm, Refuse`.
The analysis of these dialogs highlighted:
- the breaking down of the dialogs into sub-dialogs represented by plans;
- the establishment of a common ground, thanks to rewordings, agreements, questions;

- a list of speech acts, classified according to their illocutionary force and their content;
- a classification of some of these acts as positive or negative « grounding » acts;
- accommodation effects on the user.

## 5 MODELING A CONVERSATIONAL AGENT

From the corpus analysis, our aim is to design a software agent able to converse with a user and help him to find information.

### 5.1 Agent Architecture

Our agent (Figure 2) is composed of 3 main modules:
- the language model, which receives the user's inquiry in natural language. It performs a lexical and syntactical analysis (using TreeTagger (Schmid, 1994) from Stuttgart University), a pragmatic analysis (from our speech act analyzer, which uses linguistic tags — like tense, modality and context — to assign speech acts to utterances, thanks to a set of rules) and a semantic analysis (identification of terms from the CISMeF terminology).
- the dialog model, which comprises the dialog manager and the sentence generator based on incomplete sentences.
- the task model, which encapsulates the CISMeF interface to access the medical document base. It includes also a query builder from the recognized terms and a result interpreter.

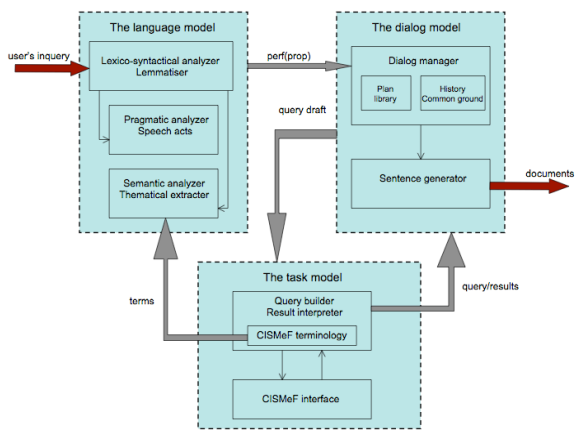

Figure 2: Conversational agent architecture

This agent is under development in Java. Our dialog uses the implementation of GoDiS (Larsson, 2002) written in Prolog. We only describe here the dialog manager.

### 5.2 Dialog Manager

The GoDiS system (Larsson, 2002) is well adapted to our needs, since it is based on an explicit task and requires no reasoning on users intention. However, it uses a list of speech acts, which is less extensive than ours: it misses acts like `Inform`, `Offer` and `Suggest`. These acts allow the system to propose relevant information in an opportunistic way according to the search.

#### 5.2.1 Overview

Our dialog manager performs a set of plans to produce speech acts. There exist two types of plans:
- question plans (planQ), in the sense of QUD, which aim at answering inquiries by returning data;
- action plans (`planA`), which run a sequence of actions.

The formalism uses the predicate logic with the operator "?" to represent questions. There are three types of questions:
- the total inquiries: ?P,
- the partial inquiries: ?P(x),
- the inquiries with a list of choices: ?set(P1(x), P2(y), P3(z)).

Moreover, our dialog manager controls an information state (IS) composed of a private part and a public part.

The private part contains:
- `Agenda`, actions of the current plan,
- `Bel`, the knowledge of the system,
- `Plan`, the current plan,
- `Nextmove`, the next speech act to be produced.
- The public part is the conversational board:
- Com, shared knowledge,
- Issue, planQ in progress or idle,
- `Qud`, focus on `Issue`,
- `Action`, `planA` in progress or idle.

Plans use a list of actions that can produce speech acts. This list comes partly from GoDiS:
- `findout(Q)` to question with the speech act `Ask`. The system repeats the question `Q` until it is answered or aborted.
- `raise(Q)` to question (only one time) optionally.

- `bind(Q)` to answer the question `Q` without posing the query.
- `assume(B)` to add a predicate `B` to the knowledge `Bel`.
- `assumeAction(A)` to add a predicate `A` to `Agenda`.
- `assumeIssue(I)` to add a predicate `I` to `Issue`.
- `consultDB(Q)` to interrogate the data base and to add relevant information to `Bel` to make suggestions.
- `cooperativeSearch(p,l,r)` to suggest to the user information having a property `p` among a list `l` in com. `r` is the result of the search (`failure` or `success`).
- `report(I)` to say the speech act `inform`.
- `say(l)` to say a speech act `l`,
- `loadPlan(p)` to load a plan `p` to be performed.
- The predicate `PostCond(P,A)` allows to give the value `A` to the predicate `P`.

Suggestions can interrupt these plans in an opportunistic way. A rule base generates them according to the IS. There exist three types of rules:
- rules to update private or shared beliefs in the IS,
- rules to choose a speech act according to the utterance just pronounced by the user,
- strategies or meta-rules to choose the update rules to be used during interactions: to update the IS with the contents of the speech act, to load plans from the plan library to `Plan`, to use accomodation rules when a non expected speech act is found, to move the current action from `Plan` to `Agenda`, to clean the IS, to perform the action in `Agenda`.

Each sub-dialog (Figure 1) is represented by a dialog plan (`PlanQ` or `PlanA`). We describe below six of them.

### 5.2.2 `Opening` Plan

The `Opening` plan allows the system to initiate the dialog with a prompt. Then the `QueryAnalysis` plan is loaded.

```
PlanA
(Opening,
  (say(Greet),
   loadPlan(QueryAnalysis)))
```

### 5.2.3 `QueryAnalysis` Plan

The `QueryAnalysis` plan aims at gathering the query of the user. If the user does not ask quickly his question, the action `Findout` allows the system to ask for his goal (definitions, documents or explanations about the system).

```
PlanA
(QueryAnalysis,
  (raise(?question(q)),
   ifThen(not q)
     findout(?set(question(Definition)),
             (question(Document)),
             (question(Explanation)))
   ifThen(question(Definition))
     loadPlan(DefinitionSearch),
   ifThen(question(Document))
     loadPlan(DocumentSearch),
   ifThen(question(Explanation))
     loadPlan(ExplanationSearch)))
```

When the user opens a dialog with the system and submits directly his query (e.g. "Hello, I would like to know if …") in one sentence, an accommodation rule allows the system to load two plans successively (`Opening` and `QueryAnalysis` plans) to adapt itself to this single sentence.

### 5.2.4 `DocumentSearch` plan

The `DocumentSearch` plan performs several steps of the sub-dialog: it builds the query and submits it to the database. Then, it evaluates the resulting documents if any.

This plan is special since it remains active in the IS. The search can be refined to increase the number of results or expanded to decrease the number of results. This plan ends only with an agreement of the user (with or without success).

```
PlanA
(DocumentSearch,
  (findout(?term(t)),
   ifThen(t)
     loadPlan(QueryBuilding(t)),
   ifThen(∃ d ∈ Bel)
     loadPlan(ListEvaluation(d))))
Post-condition: this plan remains active.
```

### 5.2.5 `QueryBuilding` plan

The `QueryBuilding` plan includes four different steps:
1. At the beginning of the search, from the initial query, the system suggests keywords of the CISMeF taxonomy thanks to the action `CooperativeAction`.
2. If the keywords found in the previous step are not sufficient to find documents, the system tries to refine the query by suggesting meta-terms and subheadings. If it does not find any term, it can ask to the user.
3. if not enough documents are found, the system expands the query,
4. if too many documents are found, the system refines the query.

```
PlanQ
(QueryBuilding(d),
 (ifThen(not ∃ keyword(k) ∈ Com)
(cooperativeSearch(keyword(k),term(t),r)
    report(submitQuery),consultDB(d)),
    ifThen(∃ keyword(k) ∈ Com
           and NotEnoughDocument ∉ Com)
      (report(refine),
       cooperativeSearch(metaTerm(m),
                         term(t),r),
      ifThen(not ∃ metaTerm(m) ∈ Com)
         raise(?metaTerm(m)),
      ifThen(not ∃ subheading(q) ∈ Com)
         raise(?subheading(q))
      report(submitQuery),consultDB(d)),
      ifThen(NotEnoughDocument ∈ Com)
     (cooperativeSearch(SpecificTerm(s),
                        term(t),r)
      ifThen(r=failure)
        (findout(?term(t)),consultDB(d)))
    ifThen(NotEnoughDocument ∉ Com)
      (report(refine),
       cooperativeSearch(SpecificTerm(s),
                         term(t),r)
      raise(?term(m)),
      ifThenElse(∃ term(t) ∈ Com)
         (consultDB(d),
          findout(?term(t)),consultDB(d)))))
```

The action `CooperativeAction` determines how to specify the inquiry to obtain relevant documents: add or delete terms, use synonyms, hyponyms, hyperonyms, etc.

### 5.2.6 `ListEvaluation` plan

The `ListEvaluation` plan takes as input a set of documents `d` and informs (as output) the user whether the documents are numerous enough or not according to the limit δ (min and max). If they are sufficient, the plan loads the plan `DocumentDescription`.

```
PlanQ
(ListEvaluation(d)
  (getNbDocuments(d,nb),
   report(nbdocuments(nb)),
   ifThen(nb<δ_min, (assume(notEnoughDocument),
          report(notEnoughDocument)))
   else(ifThen(nb>δ_max,
          (assume(tooMuchDocuments),
           report(tooMuchDocuments)))
     else(assume_issue
           (DocumentDescription(d)))))
```

### 5.2.7 `DocumentDescription` plan

The `DocumentDescription` plan takes as input a set of documents `d`, analyses their headers to decide whether they are relevant to the user's question. If necessary, the user is also given a chance to assess the relevance of the documents.

Suggestions can interrupt these plans in an opportunistic way and trigger for example a plan that explains the system. These suggestions are generated by a set of rules according to the IS.

```
PlanQ
(DocumentDescription(d),
  While(not interesting(x))
    (member(d,x),
     Report(description(x)),
     cooperativeAction(interesting(x))
     bind(?interesting(x))
    ifThen(interesting(x))
      raise(?EndOfSearch)))
```

## 6  CONCLUSION

We adopted an interdisciplinary approach to design a human-computer dialog system for health information search. We collected and analyzed a rich textual corpus on which the building of a common ground and accommodation effects on the user have been observed. Dialogs can be divided into sub-dialogs, directly linked to the task. This analysis allowed us to propose a cognitive model based on the theories of "grounding" and "accommodation". Once implemented, our system will be tested with users on the web to obtain human-computer dialogs, in order to identify and fix its shortcomings.